\documentclass[10pt,prc,aps]{revtex4}
\draft
\usepackage{graphicx}
\usepackage{color}
\usepackage{mathtools}
\begin{document}

\title{The equation of state of neutron-rich matter at fourth order of chiral effective field theory and the radius of a medium-mass neutron star}       
\author{            
Francesca Sammarruca\footnote{Corresponding author's email address: fsammarr@uidaho.edu} }                                                         
\affiliation{ Physics Department, University of Idaho, Moscow, ID 83844-0903, U.S.A. 
}
\author{            
Randy Millerson }                                                           
\affiliation{ ASML, Chandler, Arizona 85224, U.S.A. 
}
\date{\today} 

\begin{abstract}

\begin{center}
\sc {Published in {\it Universe} {\bf 2022}, {\it 8}, 133 \\
Special Issue ``Gravitational Waves as a New Probe for Astronomy and Fundamental Physics" }
\end{center}

We report neutron star predictions based on our most recent equations of state. These are derived from chiral effective field theory, which allows for a systematic development of nuclear forces, order by order. We utilize high-quality two-nucleon interactions and include all three-nucleon forces up to fourth order in the chiral expansion.
Our {\it ab initio} predictions are restricted to the domain of applicability of chiral effective field theory. However, stellar matter in the interior of neutron stars can be up to several times denser than normal nuclear matter at saturation, and its composition is essentially unknown. Following established practices, we extend our microscopic predictions to higher densities matching piecewise polytropes. The radius of the average-size neutron star, about 1.4 solar masses, is sensitive to the pressure at normal densities, and thus it is suitable to constrain {\it ab initio} theories of the equation of state. For this reason, we focus on the radius of medium-mass stars. We compare our results with other theoretical predictions and recent constraints.
\end{abstract}
%\keyword{}
\maketitle

\section{Introduction} 
\label{Intro} 
The equation of state (EoS) of neutron-rich matter is at the forefront of nuclear astrophysics because of its role in shaping the properties of neutron stars. Recently, interest in compact stars has increased considerably as we have entered the ``multi-messenger era'' of astrophysical observations. The recent GW170817 neutron star merger event has yielded new and independent constraints on the radius of the canonical mass neutron star~\cite{A_etal_1_2017, A_etal_2_2017}. Astronomy with gravitational waves provides additional opportunities to explore these exotic systems and other yet unknown regimes in the Cosmos.

The structure of a neutron star probes a very large range of densities, from the density of iron up to several times the nuclear matter saturation density, and thus no theory of hadrons can be considered reliable if extended to those regions. On the other hand, contemporary {\it ab initio} theories of nuclear and neutron matter at normal densities can be taken as the baseline for any extension or extrapolation method, which will unavoidably involve phenomenology. We recall that the radius of a 1.4 $M_{\odot}$ is sensitive to the pressure at normal densities, see Ref.~\cite{SM19}, and thus it is a suitable constraint for microscopic theories of the EoS at those densities where they are reliable.

Chiral effective field theory (EFT)~\cite{Wei92,Wei90} provides a path to a consistent development of nuclear forces. Symmetries relevant to low-energy QCD are incorporated in the theory, in particular chiral symmetry. Thus, although the degrees of freedom are pions and nucleons instead of quarks and gluons, there exists a solid connection with the fundamental theory of strong interactions through its symmetries.

Chiral EFT employs a power counting scheme in which the progression of two- and many-nucleon forces is constructed following a clear and systematic hierarchy. This allows for the inclusion of all three-nucleon forces (3NFs) which appear at a given order, thus eliminating the inconsistencies which are unavoidable when adopting meson-theoretic or phenomenological forces. Finally, it provides a clear method for controlling the truncation error on an order-by-order basis.  In the simplest approach, the latter can be expressed as the difference between the quantity computed at a given order of the chiral expansion and the one obtained at the next order. We will revisit this point in Section~\ref{error}.

We will report neutron star predictions based on our most recent EoS, which includes all subleading 3NFs ~\cite{SM21,SM_2_21}. For the reasons stated above regarding the limitations of chiral EFT, we focus on the average-mass neutron star, rather than the maximum mass of a sequence, as the latter is much more sensitive to the polytropic extensions we perform in order to complete the EoS. We also discuss the proton fraction present in our $\beta$-stable EoS, because of its relevance in neutron star cooling. Our findings are among the numerous microscopic predictions in disagreement with the outcome of PREX II and its implications for neutron stars~\cite{prexII}.

We also take the opportunity to present a succinct yet fairly complete review of the various steps in our calculations,  as well as to include some general background on neutron stars to provide context.
The paper consists of the following sections. In Section~\ref{ns_gen}, we review general aspects of neutron stars. We then outline the main aspects of our calculations, see Section~\ref{calc}. Predictions are presented and discussed in Section~\ref{res}. We summarize and conclude in Section~\ref{Concl}.

\section{General aspects of neutron stars} 
\label{ns_gen} 
A neutron star is the remnant collapsed core of a giant star which has undergone a supernova explosion. Only stars with sufficient mass, estimated to be between 8 and 25 $M_{\odot}$, undergo a supernova event at the end of their life cycle~\cite{B_1_2013}.  Due to its extremely compact nature, the neutron star is directly supported against further gravitational collapse into a black hole by mechanisms of nuclear origin, which make these objects excellent natural laboratories for exploring the nuclear EoS.

We begin by briefly summarizing the relevance of neutron stars for nuclear physics, from a historical perspective.
In 1934, just two years after the discovery of the neutron~\cite{C_1_1932}, Baade and Zwicky hypothesized the existence of a very dense stellar object, which they named neutron star, arising from the remnants of a supernova~\cite{BZ_1_1934, BZ_2_1934}. In 1939 Tolmann~\cite{T_1_1939}, and simultaneously but independently, Oppenheimer and Volkoff~\cite{OV_1_1939} estimated the mass--radius relationship of these neutron stars based on general relativity and crude nuclear force models, thus producing the famous Tolmann--Oppenheimer--Volkoff (TOV) equation. The TOV equation allows for the calculation of a theoretical upper limit on the possible mass of neutron stars. However, due to the lack of understanding of nucleonic interactions at the time, their original predictions were not accurate, placing the upper limit of a neutron star mass lower than the Chandrasekhar limit.

Over the years, with a better understanding of nuclear interactions, a more realistic picture of neutron stars and their structure emerged~\cite{C_1_1959, AS_1_1960, S_1_1960, IK_1_1965, TC_1_1966, M_1_1971, S_1_1972, SS_1_1972, KN_1_1986, IK_1_1969}.

While the existence of neutron stars was a theoretical possibility, finding proof of their existence remained a challenge. Initial efforts involving attempts to compute and observe the thermal signature of neutron stars~\cite{C_1_1964, TC_1_1966, YLS_1_1999} were unsuccessful. In 1967, Pacini~\cite{P_1_1967} postulated that fast rotating neutron stars could produce large electromagnetic emission generated from a powerful magnetic dipole. The next year, Bell and Hewish~\cite{HBPSC_1_1968} discovered the first radio pulsar, characterized by a remarkably stable periodic electromagnetic signal. Later that year, Gold theorized that neutron stellar objects were excellent candidates to explain the unusual characteristics of the pulsar signal~\cite{G_1_1968}.

By 1969, the connection between supernovas and pulsars was firmly established with the discoveries of the Vela~\cite{LVM_1_1968} and Crab Nebula~\cite{RBS_1_1969} pulsars. Hundreds of pulsars were discovered in the 1970s and 1980s using radio astronomy, while more recent developments have identified pulsars whose signals span the electromagnetic spectrum~\cite{G_1_1996}. To date, more than two and a half thousand pulsars have been discovered~\cite{M_1_2017}, these stellar objects being found in many configurations, such as binary pulsar systems~\cite{HT_1_1975}, main-sequence binary-pulsar systems~\cite{JMLBKQA_1_1992}, globular clusters~\cite{LBMKBC_1_1987}, with orbiting exo-planets~\cite{WF_1_1992}, and displaying a wide variety of unusual, yet periodic, signals~\cite{M_etal_1_2006, PTH_1_1995}. The recent GW170817 neutron star merger event, as detected through gravitational wave signatures by LIGO/Virgo~\cite{A_etal_1_2017} along with the accompanying gamma-ray burst~\cite{A_etal_2_2017}, has generated additional and remarkable observational data.

The mass--radius relationship of neutron stars is uniquely determined from the star's EoS and thus reliable observational constraints can shed light on the EoS. While the radius cannot be directly measured, the mass of neutron stars in binary systems can be inferred from observation together with the application of gravitational theory. With constraints on the mass of a star, the Doppler shift is one way to estimate the radius~\cite{G_1_1996}.

The total mass range deduced from observed neutron stars is around 1--2 $M_{\odot}$. To date, the smallest mass neutron star has been determined to be $\approx$1.17 $M_{\odot}$~\cite{M_etal_1_2015}, while the most massive observed neutron star is $\approx$2.14 $M_{\odot}$~\cite{CF_1_2020}. Of particular interest is the Chandrasekhar mass limit of white dwarf stars which is 1.4 $M_{\odot}$. If this mass is exceeded, electron degeneracy would no longer be able to support a white dwarf star from gravitational collapse. Observational constraints on neutron star masses yield values clustered around 1.4 $M_{\odot}$~\cite{Z_etal_1_2011} and for this reason this value is taken as the neutron star canonical mass. This also led to the idea that white dwarf collapse may be an additional  mechanism for the formation of neutron stars~\cite{TSYL_1_2013, SQB_1_2015, RFBSCK_1_2019}.

The neutron star radius is not measured directly, but observational data allow for indirect inference. Observation-based constraints consistently place the estimated radius of a neutron star in the range of 10--15 km. For instance, using accreting and bursting sources, the radius of the canonical-mass neutron star was determined to be within a range of 10.4 to 12.9 km~\cite{SLB_1_2013}, while analysis from the LIGO/Virgo observations determined the radius to be between 11.1 and 13.4 km~\cite{AGKV_1_2018}. Upper limits on the neutron star radii, as determined from iron emission lines, were placed between 14.5 and 16.5 km~\cite{C_etal_1_2008}.

Neutron star models are generally in good agreement with observational constraints for the radius. For instance, the radius of the canonical-mass neutron star predicted  from the set of EoS applied in Ref.~\cite{LS_1_2014} is predicted to be in the range 10.45--12.66 km. From a variety of techniques, based on experimentally determined quantities correlated to symmetry energy parameters, the radius is determined to be between 10.7 and 13.1 km~\cite{NGHL_1_2011, T_etal_1_2012, LL_1_2013, LS_1_2014}, while using a range of theoretical models a limit of 9.7 to 13.9 km is obtained~\cite{SG_1_2012, HLPS_2_2013, LS_1_2014}. Recent surveys of neutron star physics and theoretical approaches include Refs.~\cite{OHKT_1_2017, BV_1_2020, V_1_2018}. Exotic matter in stars is addressed, for instance, in Ref.~\cite{BBS_1_2016}.

On theoretical grounds, the largest mass was predicted to be 3.2 $M_{\odot}$~\cite{RR_1_1974}, based on only three assumptions: (1) General Relativity is the appropriate theory to describe these massive stars;  (2) the EoS is constrained by Le Chatelier's principle ($\partial P$/$\partial \epsilon \geq$ 0);  (3) the causality condition, which constrains the speed of sound in dense matter to remain below the speed of light. While such massive neutron star may be theoretically possible, none has been observed in this mass range.

It is interesting to note the small range of values for the radius across the mass range of neutron stars.  Heavier neutron stars have larger central densities and thus the star becomes comparatively more compact, resulting in a very narrow mass--radius range,  in contrast to main-sequence stars whose masses and radii span several orders of magnitude.

\section{Description of the calculation} 
\label{calc}
The EoS for neutron and symmetric matter are obtained at the leading-order in the hole-line  expansion---namely, {via} a non-perturbative calculation of the particle--particle ladder.  As pointed out in Refs.~\cite{SM21,SM_2_21}, the third-order hole-hole diagram was considered and found to be very small at normal density~\cite{Cor14}. In comparison, the particle-hole diagram is larger, but its contribution is still relatively minor---we estimate an uncertainty in the order of $\pm$1 MeV on the potential energy per particle at saturation.
The single-particle potentials are computed self-consistently with the $G$-matrix, employing a continuous spectrum.

Next, we proceed to describe the input two-nucleon forces (2NFs) and 3NFs.

\subsection{The Two-Nucleon Force}
\label{II}

The 2NF we apply are from Ref.~\cite{EMN17}, a family of high-quality potentials from leading order (LO) to fifth order (N$^4$LO) of the chiral EFT. At fifth order, the nucleon-nucleon ($NN$) data below pion production threshold are reproduced with the excellent precision of $\chi ^2$/datum = 1.15.

The interactions in this set are more internally consistent than those from the previous generation~\cite{EM03}. Furthermore, the long-range part of these potentials is tightly constrained by the $\pi N$ low-energy constants (LECs) from the Roy--Steiner analysis of Ref.~\cite{Hofe+}. This analysis is sufficiently accurate to render errors in the $\pi N$ LECs essentially negligible for the purpose of quantifying the uncertainty.

To suppress high-momentum components in the potential when solving the non-perturbative Lippmann-Schwinger equation, one must apply a regulator, for which we choose the non-local form:
\begin{equation}
f(p',p) = \exp[-(p'/\Lambda)^{2n} - (p/\Lambda)^{2n}] \, ,
\label{reg}
\end{equation}
expressed in terms of the final  ($p' \equiv |{\vec p}\,'|$) and initial ($p \equiv |\vec p \, |$) nucleon momenta in their center-of-mass system. With the choice
$\Lambda$ = 450 MeV---which we maintain throughout this paper---the potentials are soft according to the
Weinberg eigenvalue analysis of Ref.~\cite{Hop17} and the perturbative calculations of infinite matter from Ref.~\cite{DHS19}.

\subsection{The Three-Nucleon Force}
\label{III}

In the framework of the $\Delta$-less chiral EFT (which we apply),
the first occurrence of 3NF is seen at the third order. The leading 3NF consists of three components~\cite{Epe02}: the long-range two-pion-exchange (2PE) graph, which depends on LECs  $c_1, c_3$, and $c_4$, the medium-range one-pion-exchange (1PE) diagram, carrying the LEC $c_D$, and a short-range contact term, containing the LEC $c_E$. We recall that the terms depending on $c_4$, $c_D$, and $c_E$ do not contribute in neutron matter~\cite{HS10}.

In infinite matter, it is possible to construct approximate expressions for the 3NF as density-dependent effective two-nucleon interactions as derived in Refs.~\cite{holt09,holt10}. These can be written in terms of the well-known non-relativistic two-body nuclear force operators and, thus, can be easily implemented in the $NN$ partial wave formalism for the $G$-matrix, which leads to the EoS.
One must be careful to avoid overcounting in the 3NF when the latter is represented as  density-dependent potentials, which is accomplished by including the appropriate combinatoric factor in the calculation of the energy per particle, as done in Ref.~\cite{SM21,SM_2_21}.

The topologies of the effective density-dependent two-nucleon interactions originate from the corresponding chiral 3NF. At N$^2$LO, there are six one-loop topologies. Of those, three  come from the 2PE graph of the chiral 3NF and contain the LECs $c_{1,3,4}$, which also appear in the 2PE part of the $NN$ interaction. Of the remaining three one-loop topologies, two originate from the 1PE diagram of the 3NF, and depend on the LEC $c_D$. The last one-loop diagram is the 3NF contact term, with LEC $c_E$.

We also include the subleading (N$^3$LO) 3NF, derived in Ref.~\cite{Ber08,Ber11}. References~\cite{Tew13,Dri16,DHS19,Heb15a} report applications of the subleading 3NF in some many-body systems. 
The long-range part of the 3NF at N$^3$LO includes the 2PE topology, which is the longest-range contribution, the two-pion-one-pion exchange (2P1PE) topology, and the ring topology, which represents a pion being absorbed and reemitted from each of the three nucleons.
Again, in-medium $NN$ potentials can be obtained from these topologies. For the long-range subleading 3NFs, the expressions are given in Ref.~\cite{Kais19} for symmetric nuclear matter (SNM) and in Ref.~\cite{Kais20} for neutron matter (NM). The short-range subleading 3NF includes the following topologies: the one-pion-exchange-contact (1P-contact), which ends up giving a vanishing net contribution, the two-pion-exchange-contact (2P-contact), and relativistic corrections, which depend on the $C_S$ and the $C_T$ LECs of the 2NF and are proportional to $1/M$, where $M$ is the nucleon mass. We include relativistic corrections as well and find them to be very small (less than one MeV). The expressions for the in-medium $NN$ potentials generated by the short-range subleading 3NFs are taken from Ref.~\cite{Kais18} for SNM and Ref.~\cite{Treur} for NM.

Table~\ref{tab1} displays the LECs we use, which are taken  from Ref.~\cite{DHS19}.  An important point to highlight: the largest part of the subleading two-pion-exchange 3NF has a very similar formal structure to the leading 2PE~\cite{KGE12} and thus a large part of the subleading two-pion-exchange 3NF can be effectively accounted for with a shift of the LECs. For this reason, when the subleading 3NFs are included, we replace the $c_1$, $c_3$, and $c_4$ LECs shown in Table~\ref{tab1} with $-$1.20 GeV$^{-1}$, $-$4.43 GeV$^{-1}$, and 2.67 GeV$^{-1}$, respectively, corresponding to  shift of $-$0.13 GeV$^{-1}$ (for $c_1$), 0.89 GeV$^{-1}$ (for $c_3$), and $-$0.89 GeV$^{-1}$ (for $c_4$)~\cite{Ber08}.
 
\begin{table*}[t]
\caption{
Values of the LECs $c_{1,3,4}$, $c_D$, $c_E$, $C_S$, and $C_T$ at N$^2$LO and N$^3$LO.
The momentum-space
cutoff $\Lambda$ is equal to 450 MeV.
The LECs $c_{1,3,4}$ are given in units of GeV$^{-1}$, while
$c_D$ and $c_E$ are dimensionless.}
\label{tab1}
\begin{tabular*}{\textwidth}{@{\extracolsep{\fill}}ccccccccc}
%\toprule
\hline
\hline
& \boldmath{$\Lambda$}\textbf{ (MeV)} & \boldmath{$c_1$} & \boldmath{$c_3$ }& \boldmath{$c_4$} &  \boldmath{$c_D$} &\boldmath{ $c_E$}  &\boldmath{ $C_S$}  & \boldmath{$C_T$} \\
%\midrule
\hline
\hline
N$^2$LO & 450 &  $-$0.74 & $-$3.61 & 2.44  &   2.75 &   0.13  & $-$0.013000 & $-$0.000283 \\
%\midrule
\hline
N$^3$LO & 450 & $-$1.07 & $-$5.32 & 3.56  &  0.50 &   $-$1.25  &$-$0.011828 &  $-$0.000010 \\
%\bottomrule
\hline
\hline 
\end{tabular*}
\end{table*}

\begin{figure*}[t]
\centering
\hspace*{-1cm}
\includegraphics[width=7.5cm]{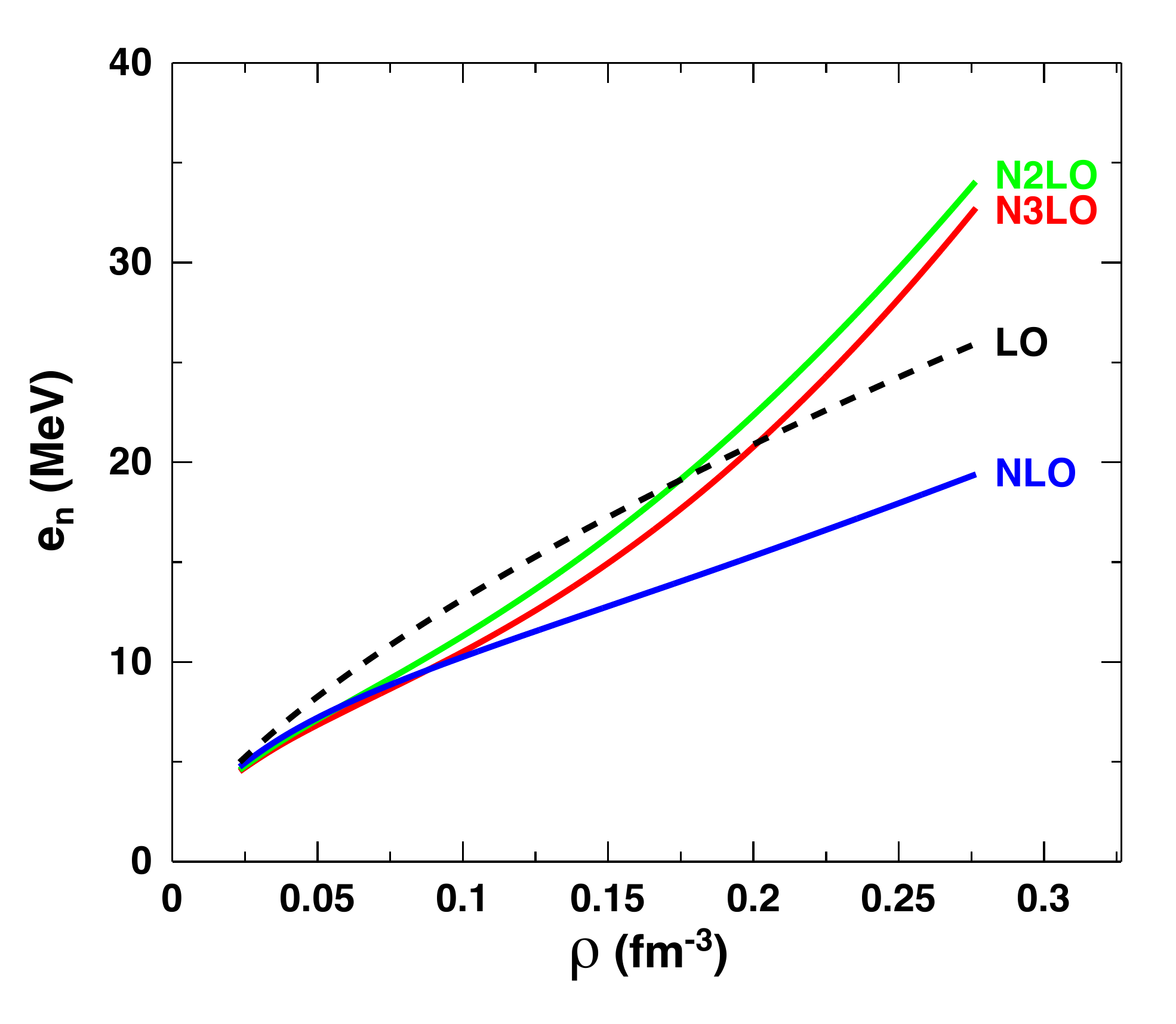}\hspace{0.01in}
\vspace*{0.05cm}
\caption{ Energy per particle in neutron matter as a function of density from leading to fourth order of chiral perturbation theory. The cutoff is fixed at 450 MeV. The EoS are those obtained in Ref.~\cite{SM21}.
}
\label{obo_nm}
\end{figure*}

In Figure~\ref{obo_nm}, we show the EoS in NM over four orders, from LO to N$^3$LO~\cite{SM21}.  Large variations at low orders are of course not surprising, nor is the remarkable impact of the leading 3NF at N$^2$LO. The transition to fourth order brings in a slight increase in attraction, as was found from other EFT-based predictions~\cite{Dri16}. The overall convergence pattern is encouraging.

\begin{figure*}[t]
\centering
\hspace*{-1cm}
\includegraphics[width=7.5cm]{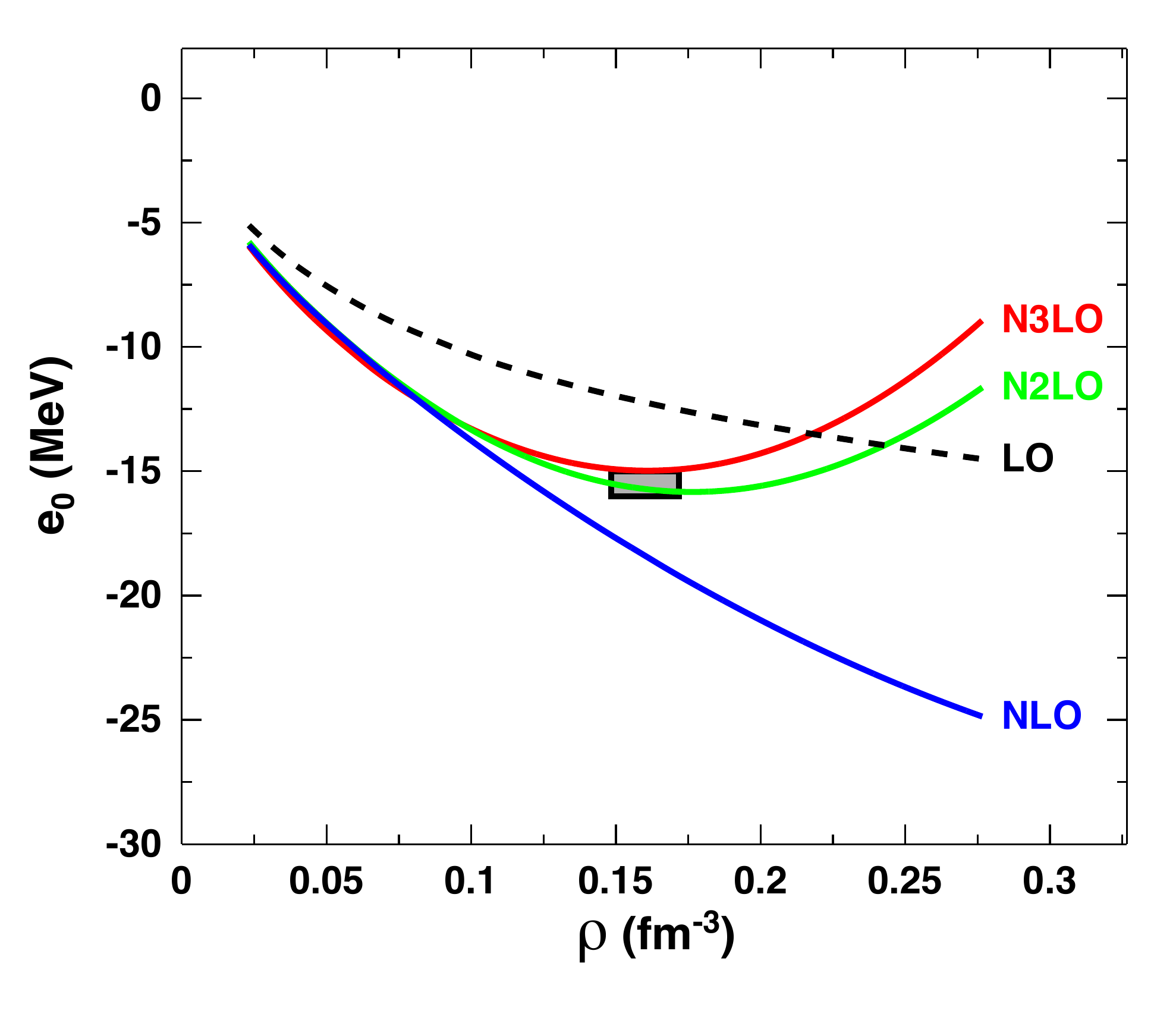}\hspace{0.01in}
\vspace*{0.05cm}
\caption{ Energy per particle in symmetric nuclear matter as a function of density from leading to fourth order of chiral perturbation theory. The cutoff is fixed at 450 MeV. The shaded box marks the empirical saturation point. The EoS are the same as obtained in Ref.~\cite{SM_2_21}. %%Please change hyphen to minus. %%I don't understand this comment%
}
\label{obo_c}
\end{figure*}

In Figure~\ref{obo_c}, an analogous presentation is provided for SNM. Similar considerations apply with regards to the order-by-order convergence pattern and the 3NF ``signature'' as in~NM.

\subsection{Equation of State for Stellar Matter}
\label{beta}

In this section we outline the main steps to construct the EoS for stellar matter in $\beta$-equilibrium. We define the total energy per baryon as:
%Eq. 1
\begin{equation}
\label{total_eng_per_prt}
\begin{aligned}
e_T (\rho) = e_0(\rho) + e_{sym} \ (Y_{n}-Y_{p})^2 +
e_e + e_\mu + \sum_{i = n,p} Y_i m_i  \; ,
\end{aligned}
\end{equation}
where $Y_{n,p}$ stands for the neutron or proton fraction. The last term accounts for the baryon rest energy, while $e_{e/\mu}$ are the energies (per baryon) of the electrons and muons, respectively. All terms are functions of density.

The energy density ($\epsilon_{i}$), pressure ($p_{i}$), and number density ($\rho_{i}$) for each particle species, $i$, at a given Fermi momentum, $(k_{F})_{i}$, can be expressed as:
\begin{equation}
\label{eng_den}
\epsilon_i = \frac{\gamma}{2\pi^2} \int^{k_{F_{i}}}_{0} \sqrt{k^2 + m_{i}^{2}} \ k^2 \ dk  \; ,
\end{equation}
\begin{equation}
\label{prs}
p_i = \frac{1}{3} \frac{\gamma}{2\pi^2} \int^{k_{F_{i}}}_{0} \frac{k^2}{\sqrt{k^2 + m^2_{i}}} \ k^2 \ dk  \; ,
\end{equation}
\begin{equation}
\label{den}
\rho_i = \frac{\gamma}{2\pi^2} \int^{k_{F_{i}}}_{0} k^2 \ dk  \; ,
\end{equation}
where $\gamma$ is the spin/isospin degeneracy factor. The energy per volume is obviously related to the energy per particle:
\begin{equation}
\label{eps_eq_den_der}
\epsilon = \rho \; e(\rho)  \; .
\end{equation}%Please confirm if it need indent, same in the following highlight.
The Fermi energy for each species, $i$, or chemical potential, is given by
\begin{equation}
\label{chempot}
\mu_{i} = \frac{\partial \epsilon_{i}}{\partial \rho_{i}} = \sqrt{k^{2}_{F_i}+m_{i}^{2}}  \; .
\end{equation}
The fractions of each particle species,
\begin{equation}
\label{frac_expr}
Y_i=\frac{\rho_i}{\rho}  \; ,
\end{equation}
is related to the chemical potential through:
\begin{equation}
\label{chempot_frac}
\mu_{i} = \frac{\partial \epsilon_{i}}{\partial \rho_{i}} = \frac{\partial e_{i}}{\partial Y_{i}}  \; .
\end{equation}
From the condition of energy minimization, subjected to the constraints of conserved nucleon density and global charge neutrality,
\begin{equation}
\label{frac_bary}
\rho_p + \rho_n = \rho \quad \mbox{or} \quad Y_p + Y_n = 1  \; ,
\end{equation}
\begin{equation}
\label{frac_chrg}
\rho_p = \rho_e + \rho_{\mu} \quad \mbox{or} \quad Y_p = Y_e + Y_{\mu}  \; ,
\end{equation}
one can obtain all particle fractions.

\subsection{Polytropic extrapolation} 
\label{poly}
Chiral predictions have a limited domain of validity, which, in the previous section, we estimated to be about twice the saturation density. The densities within neutron stars can reach five to six times saturation density, and therefore an appropriate method for extrapolating the EoS to these densities must be employed. To accomplish this, we express the high density pressure through polytropes~\cite{RLOF_1_2009}:
\begin{equation}
\label{prs_poly}
P(\rho) =  \rho^2 \frac{\partial e_T(\rho)}{\partial \rho} = \alpha \rho^{\Gamma} \; ,
\end{equation}
where $\alpha$ is chosen such as to ensure continuity at the matching density.  A comment is in place: while continuity of the pressure is of course preserved, additional considerations are necessary to ensure continuity of the derivative. The latter would be essential to implement thermodynamic consistency of the piecewise EoS, which is beyond our present scope. Note, further, that the presence of discontinuities in the polytropic index is not unusual for the purpose of describing the global features of the star~\cite{RLOF_1_2009}.
Following Ref.~\cite{SM19},  we match piecewise polytropes to the {\it ab initio} predictions as explained next.

The microscopic predictions reach a Fermi momentum of 1.6 fm$^{-1}$, which corresponds to 2.016 fm$^{-1}$ in pure neutron matter at the same density, $\rho$ = 0.277 fm$^{-3}$. The standard practice is to ensure that the characteristic momentum of the system, $p$, divided by the cutoff, $\Lambda$, is a reasonable expansion parameter. Taking $p$ to be the average momentum in a free Fermi gas of neutrons at the highest density we consider, we obtain a value of 68\% for $p/ \Lambda$, which is a pessimistic estimate, since the average momentum in $\beta$-stable matter is smaller than in pure NM. Having chosen the matching density, $\rho_1$, we join the pressure predictions with polytropes of different adiabatic index, ranging from 1.5 to 4.5. This range is chosen following guidelines from the literature, in particular Ref.~\cite{RLOF_1_2009}, where constraints on phenomenologically parameterized neutron-star equations of state are investigated. To simulate a (likely) scenario where the pressure displays different slopes in different density regimes, we define a second matching density, $\rho_2$, approximately equal to 2$\rho_1$, at which point a set of polytropes covering the same range of $\Gamma$ is attached to each of the previous polytropes, yielding a total of 49 possible combinations. This is illustrated in Figure~\ref{pgam}.
It is important to emphasize that high-density EoS extrapolations are not meant to be a replacement for microscopic theoretical predictions~\cite{SM19} which, at this time, are not feasible at super-high densities. Instead, the spreading of the high-density pressure values from the piecewise variation of the polytrope index allows to probe the sensitivity of lower-density predictions to the much larger uncertainty at high density.

To construct a physical EoS for high densities, we must apply additional constraints. One is the causality limit, which imposes the speed of sound in matter to be less than the speed of light. In terms of
$P(\epsilon)$, the causality condition reads:
\begin{equation}
\label{sos}
\frac{d P(\rho)}{d \epsilon(\rho)} < 1 \; .
\end{equation}

Note that the constraint on the speed of sound is strictly valid only in the absence of dispersion or absorption in stellar matter~\cite{W_1_1999}. 
Nevertheless, imposing the causality constraint is standard practice when constructing neutron star EoS and we will apply it 
 in this work~\cite{RR_1_1974, G_1_1996, W_1_1999, RLOF_1_2009}.

Additionally we will only consider polytropes, which can support a maximum mass of at least 2.01 $M_{\odot}$, to be consistent with the lower limit of the (2.08 $\pm$ 0.07) $M_{\odot}$ observation reported in Ref.~\cite{PSRJ0740} for the J0740 + 6620 pulsar along with a radius estimate of (\mbox{12.35 $\pm$ 0.75) km}.
To complete the EoS on the low-density side, we attach a crustal EoS~\cite{HW_1_1965, NV_1_1973}.

\subsection{Mass-radius relation} 
\label{MR}
With the EoS available over a full range of densities, we move to the mass--radius relation in a neutron star. In this section we will briefly review the relativistic equations for hydrostatic equilibrium, the TOV equations~\cite{T_1_1939, OV_1_1939} and how the mass--radius relationship emerges from them for a given input EoS.

The TOV equation describes a spherically symmetric inertial massive object composed of a perfect fluid in hydrostatic equilibrium. The equation relates the pressure within the star to the mass-energy density as functions of the radial distance from the star's center:

\begin{equation}
\label{TOV}
\frac{d P(r)}{d r} = - \frac{G}{c^{2}} \ \frac{(P(r) + \epsilon (r) ) \ (M(r) + 4 \pi r^3 \frac{P(r)}{c^{2}})}{r (r - \frac{2 G M(r)}{c^{2}})} \; .
\end{equation}
A spherical shell of material is related to the energy density at a distance $r$  from the star's center by:
\begin{equation}
\label{dMR}
\frac{d M(r)}{dr} = \frac{4 \pi}{c^{2}} r^{2} \epsilon(r) \; .
\end{equation}
The star's gravitational mass ($M$) is determined from the radius ($R$) and the mass-energy density ($\epsilon(r)$):
\begin{equation}
\label{MR}
M(R) = \int^{R}_{0} \frac{\epsilon(r)}{c^{2}} \ d^{3} r \; .
\end{equation}

Since the pressure and energy-density are functions of density, for a fixed central density the mass--radius of the star can be determined by Equations~(\ref{dMR}) and~(\ref{TOV}). Equation~(\ref{TOV}) can be integrated numerically by summing over shells of fixed width at incremented distance from the star's center so as to evaluate the total pressure as a function of radial distance. Equation~(\ref{dMR}) can be integrated in the same fashion, simultaneously, to determine the mass contained within each spherical shell. To accomplish this, we employed the fourth-order Runge--Kutta method.
The radial distance at which the pressure effectively vanishes corresponds to the star's radius. Then, Equation~(\ref{MR}) provides the total mass enclosed within such radius.

\section{Results and Discussion }
\label{res}
\subsection{Chiral Uncertainty}
\label{error}
First, some comments on the estimation of the chiral error.

As pointed out in Section~\ref{II}, errors in the $\pi N$ LECs are small enough to be neglected when quantifying the combined uncertainty. The truncation error is of course central to the philosophy and the application of chiral EFT, being the indicator of the quality of the convergence pattern.

In the most basic approach, one may argue that,
if observable $X$ has been determined at order $n$ and at order $n+1$, a reasonable estimate of the truncation error is simply the difference between the value  of $X$ at order $n$ and the one at the next order:
\begin{equation}
\Delta X_n = |X_{n+1} - X_n| \; .
\label{del}
\end{equation}

To estimate the uncertainty at the highest available order, one needs additional considerations. We follow the prescription of Ref.~\cite{Epel15}. First, one needs to identify the typical momentum or energy scale for the system being considered, say, $p$. Defining $Q$ as the largest between $\frac{p}{\Lambda_b}$ and $\frac{m_{\pi}}{\Lambda_b}$, where $\Lambda_b$ is the breakdown scale of the chiral EFT, about 600~MeV~\cite{Epel15}. The truncation error for the observable $X$ at N$^3$LO is then defined as:
\begin{equation}
\Delta X = \max \{Q^5|X_{LO}|, Q^3|X_{LO} - X_{NLO}|,Q^2|X_{NLO} - X_{N^2LO}|, Q|X_{N^2LO} - X_{N^3LO}| \} \; .
\label{err}
\end{equation}
We take $p$ to be the average momentum of a Fermi gas of neutrons, $\sqrt{\frac{3}{5}}k_F^n$.

\subsection{Results}
\label{ns_res}
\begin{figure*}[!t] 
\centering
\hspace*{-1cm}
\includegraphics[width=7.5cm]{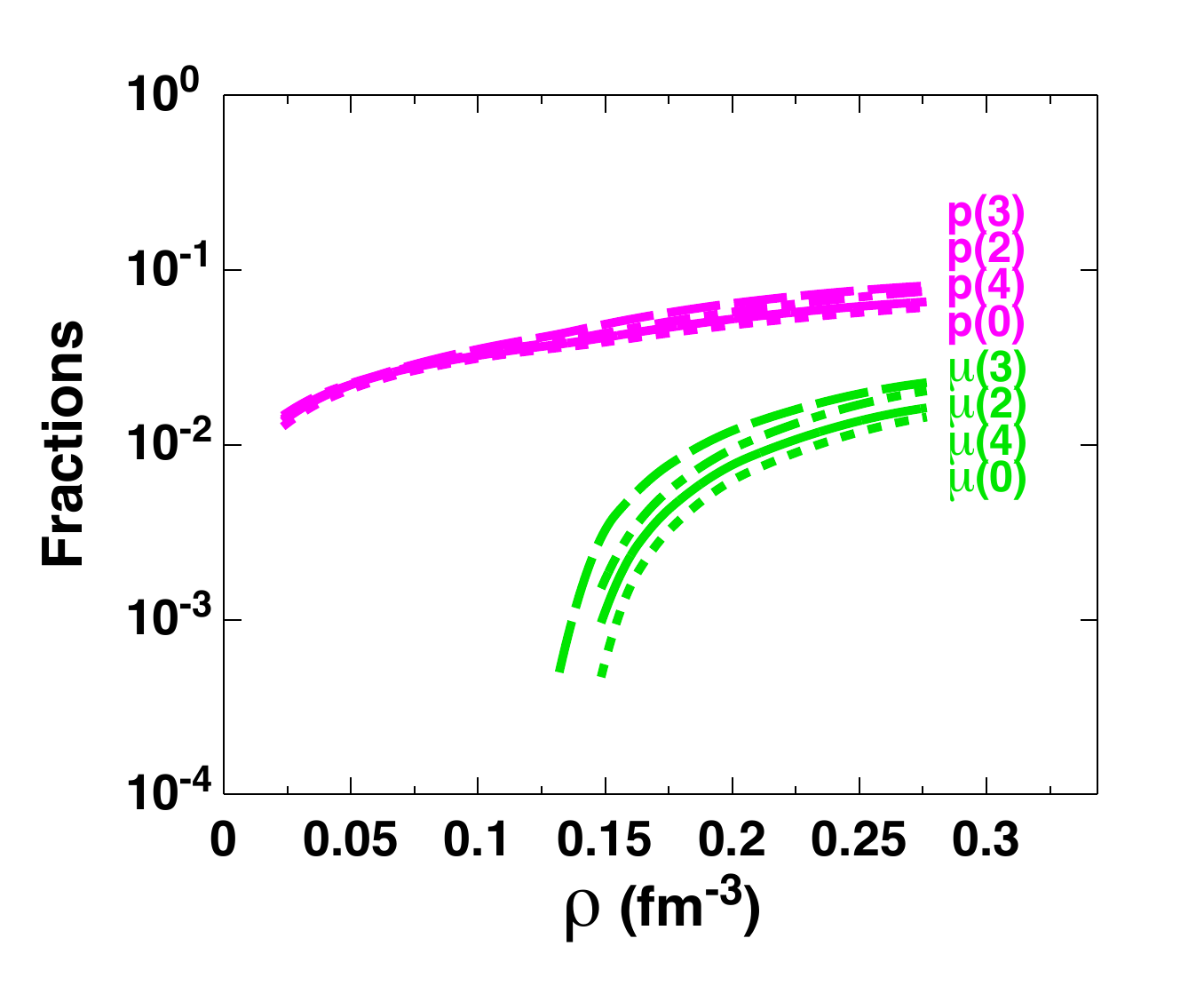}\hspace{0.01in} 
\vspace*{0.05cm}
 \caption{ Proton fraction and muon fraction as a function of density. The numbers in the curve labels (0, 2, 3, 4) indicate the chiral orders LO, NLO, N$^2$LO, and N$^3$LO, respectively. See text for more details.
}
\label{frac}
\end{figure*}

\begin{figure*}[!t] 
\centering
\hspace*{-1cm}
\includegraphics[width=7.5cm]{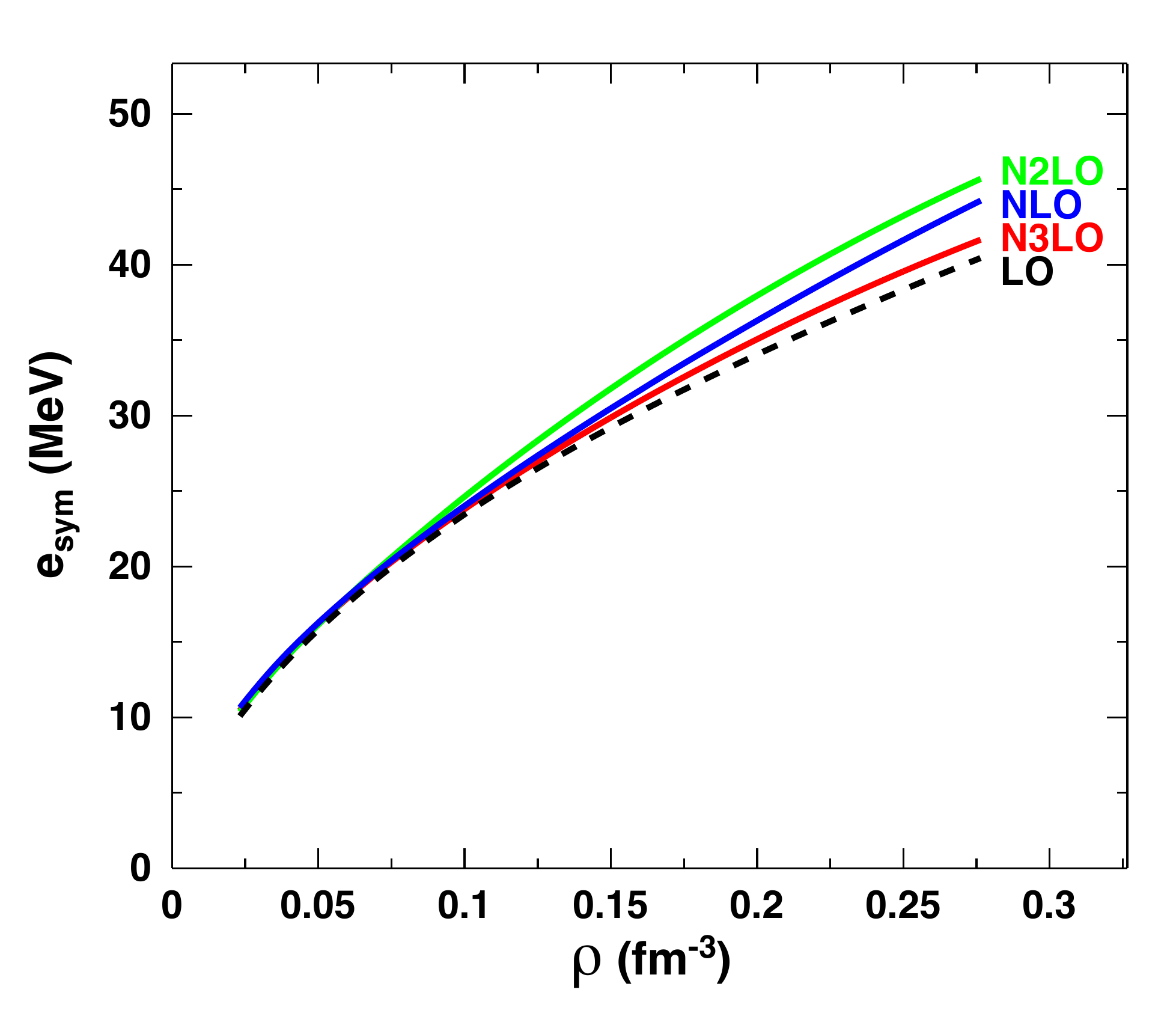}\hspace{0.01in} 
\vspace*{0.05cm}
 \caption{Symmetry energy as a function of density, order by order.
}
\label{esym}
\end{figure*}

In Figure~\ref{frac}, we show the proton fractions we obtain from leading to fourth order of chiral perturbation theory. Electron fractions are not shown to avoid excessive crowding---prior to the onset of muons, electron and proton fractions are of course equal, and remain close to each other, as it can be inferred from the small values of the muon fractions. Our proton fractions are very small, consistent with the relatively soft symmetry energy, see Figure~\ref{esym}, which brings up the issue of direct Urca processes---the most effective neutrino emission mechanism. Our predictions are far from the direct Urca threshold of approximately 11\% around normal density. Instead, we find for the N$^3$LO proton fraction a normal density a value of $0.046 \pm 0.0035$.

 To offer the reader a broader overview, we show in Table~\ref{par0} the value of the symmetry energy and the slope parameter $L$ with their uncertainty, 
where $L$ is defined as
\begin{equation}
\label{L}
L=3\rho_{o} \Big( \frac{\partial e_{sym}(\rho)}{\partial \rho} \Big)_{\rho_{o}}  \; .
\end{equation}
These values were shown in Ref.~\cite{SM21} and compared to recent constraints.

\begin{table*}
\caption{The symmetry energy and the slope parameter at N$^3$LO at saturation density $\rho_o$. 
$L$ is defined as in Eq.~(\ref{L}).}
\label{par0}
\centering
\begin{tabular*}{\textwidth}{@{\extracolsep{\fill}}ccc}
\hline
\hline
 $\rho_0(fm^{-3})$  & $e_{sym}(\rho_o)$ (MeV)  & $L (\rho_o)$(MeV)   \\
\hline
\hline
 0.16 &  31.3 $\pm$ 0.8  & 52.6 $\pm$ 4.0  \\
\hline
\hline
\end{tabular*}
\end{table*}

\begin{figure*}[!t] 
\centering
\hspace*{-1cm}
\includegraphics[width=8.5cm]{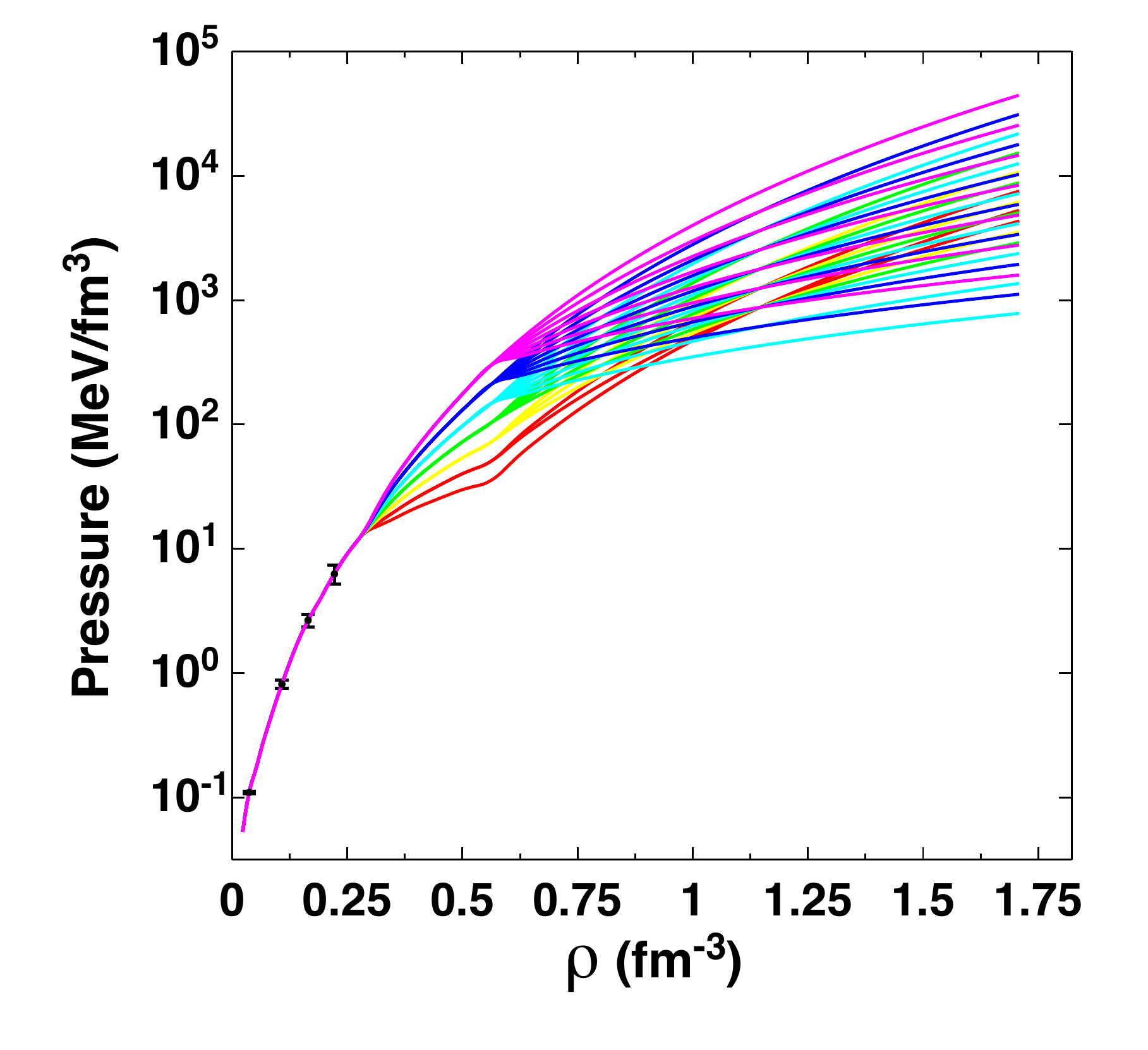}\hspace{0.01in} 
\vspace*{0.05cm}
 \caption{ Pressure in $\beta$-stable matter as a function of density.  The figure shows the spreading of the pressure values due to the matching of polytropes at two densities, $\rho_1$=0.277 fm$^{-3}$ and $\rho_2$=0.506 fm$^{-3}$. Each group of curves with the same color contains EoSs with the same $\Gamma_1$ and varying $\Gamma_2$. The microscopic predictions (single pink curve prior to the first matching point), are obtained at N$^3$LO and cutoff equal to 450 MeV. }
\label{pgam}
\end{figure*}

We proceed with pressure predictions. Figure~\ref{pgam} displays the pressure in stellar matter as a function of the number density. The various curves span the range of acceptable combinations of polytropes,
as explained in Section~\ref{poly}.
In Table~\ref{tabR} we present predictions for the radius, central density, and speed of sound for M = 1.4 $M_{\odot}$ for those combinations.
Several comments are in place.
First, we note that, overall, the radius is not very sensitive to the $\Gamma_1$, $\Gamma_2$ variations.
For values of $\Gamma_1$ on the low end of the range, acceptable combinations require values of $\Gamma_2$ on the higher end---understandable in terms of maximum mass constraints.
Further, for a fixed value of $\Gamma_1$, the sensitivity of the radius to variations in $\Gamma_2$ is essentially negligible, and of course it vanishes when the central density is below the second matching density.
One more comment regarding causality: While all EoS are causal at the central densities displayed in Table~\ref{tabR}, some may not be so at the central density of the maximum mass of the sequence. In such cases, the M(R) correlation is typically truncated at the causality limit, retaining the EoS at the lower densities.

The mean value and standard deviation are  (${\bar R}_{1.4}$ = 11.96 $\pm$ 0.58) km. The same procedure is applied at the lower orders---LO to
N$^2$LO---that is, the EoS at each order is extended with polytropes and the mean value of the radius is calculated.
With radius predictions available from leading to fourth order, we determine the truncation error {via} the prescription in Equation~(\ref{err}), where we take $p$ to be the neutron Fermi momentum at saturation density. This choice is reasonable because the {\it average} density of a neutron star is comparable to saturation density and because of the strong sensitivity of the radius to the normal density region.
From Equation~(\ref{err}) we obtain an uncertainty of  $\pm $ 0.54 km. Combining the truncation and extrapolation uncertainties in quadrature, we state our estimate of the radius as:
\begin{equation}
R_{1.4} = (11.96 \pm 0.80) \; \mbox{km} \; ,
\label{R1.4}
\end{equation}
in excellent agreement with the LIGO/Virgo range of 11.1 to 13.4 km~\cite{AGKV_1_2018}.

Our result is within the range generally found with EoS based on chiral EFT, which is 10 km to 14 km~\cite{DHW21, HLPS10}, accounting for additional theoretical uncertainties, such as those originating from the choice of the many-body method and the implementation of the 3NF~\cite{SCGF1, SCGF2,QMC1,QMC2, QMC3}. Some sensitivity of  $R_{1.4}$ to the matching density was found~\cite{LH19, TCGR18}. Moving the matching density from $\rho_0$ to 2$\rho_0$ changed the range to (9.4--12.3) km~\cite{TCGR18} and to (10.3--12.9) km~\cite{LH19}.  In the present analysis, the first matching point is determined by the highest density we reach out with the EFT calculations--a natural matching point. As for the second matching density, Table~\ref{tabR} shows that the details of the extension at the higher densities has only a minor impact on $R_{1.4}$.

\begin{table*}[t]
\caption{Radius, $R$, central density, $\rho_c$, and speed of sound at central density, $v_s/c$ (in units of the speed of light), for M = 1.4 M$_{\odot}$ at N$^{3}$LO and $\Lambda$=450 MeV. The shown values of $\Gamma_1$ and $\Gamma_2$ are the combinations of polytropic indeces consistent with both the maximum mass and the causality constraints up to the relevant densities. When the star's central density is below the second matching density, there is no dependence on $\Gamma_2$.
}
\label{tabR}
\begin{tabular*}{\textwidth}{@{\extracolsep{\fill}}ccccc}
\hline
\hline
  $\Gamma_1$ & $\Gamma_2$ & R (km) & $\rho_c$ (fm$^{-3}$) &  $v_s/c$   \\
\hline    
\hline 
  1.5 & 4.5 & 10.05  & 0.82  &   0.90  \\
\hline
  2    & 4  &  10.64   & 0.74  &   0.81  \\
     & 4.5 &  10.70  & 0.71  &   0.86  \\
\hline
 2.5  & 3.5 &  11.34  & 0.63  &   0.71 \\
        & 4    &  11.34  & 0.62  &  0.78 \\
        & 4.5 &   11.35  & 0.61 & 0.84  \\
\hline
  3   & 3  &  11.87 & 0.51  &   0.62  \\
        & 3.5  &  11.87  & 0.51  &   0.63  \\
        & 4    & 11.87    & 0.51  & 0.63  \\ 
        & 4.5 &  11.87   & 0.51  & 0.63 \\
\hline 
 3.5  & any & 12.15 &  0.45 & 0.67 \\    
\hline 
  4    & any & 12.29 & 0.41 & 0.71 \\
\hline 
  4.5    & any & 12.39  & 0.39 & 0.75  \\
\hline
\hline
\end{tabular*}
\end{table*}

Closely related to the star radius is the tidal deformability, which is a measure of how easily the star deforms under an external tidal field. Therefore, a smaller compact star will have smaller tidal deformability as compared to a larger, less dense star. The tidal deformability is defined as the ratio of the induced quadrupole $Q_{ij}$ to the perturbing tidal field $E_{ij}$, which explains the sensitivity to the radius~\cite{Poiss14}. The dimensionless tidal deformability in the mass range that is relevant for GW170817 (and for the present analysis)---$1.1M_{\odot} \leq M \leq 1.6M_{\odot}$---is found to be proportional to $\Big (\frac{M}{R} \Big )_{sc}^{-6} $, where
\begin{equation}
\Big (\frac{M}{R} \Big)_{sc} = \frac{GM}{Rc^2}
\label{beta}
\end{equation}
is the (dimensionless) star compactness. The dimensionless tidal deformability can then be expressed as:
\begin{equation}
\Lambda = a \Big (\frac{M}{R} \Big)^{-6}_{sc} \; ,
\label{tide}
\end{equation}
where $a = 0.0093 \pm 0.0007$~\cite{De+2018}.
Using our estimate for the radius as in Equation (\ref{R1.4}), we obtain from Equation~(\ref{tide}) a range of $\Lambda$ between 247 and 474, in excellent agreement with the Bayesian analysis of Ref.~\cite{LH19}, where the limits of 68\% credibility are found to be 249 and 465, with $\Lambda$ = 379 the most probable value.

Recent analyses of the neutron star radius include Refs.~\cite{Capano+2020, TMR2019}. Our calculations differ from those in several ways, most importantly, the derivation of the microscopic part of the EoS is based on local chiral interactions at N$^2$LO---by contemporary standards, high-precision description of $NN$ data requires the construction of potentials at the fourth order. Furthermore, all subleading 3NF at the same order have been available for some time. For these reasons, modern chiral EFT-based nuclear forces should be consistently at N$^3$LO.
Other differences include the many-body theory---local interactions are constructed to be used in QMC calculations---and the derivation of the $\beta$-stable EoS, which we obtain from the microscopic EoS for both NM and SNM, as opposed to considering pure neutron matter, potentially extrapolated to $\beta$-stability.

In Ref.~\cite{nicer1}, a Bayesian estimation of the mass and radius of PSRJ0030 + 0451 based on NICER data gives $M/M_{\odot} = 1.34 (+0.15, -0.16)$ and $R = 12.71 (+1.14, -1.19)$ km for the equatorial radius. The authors of Ref.~\cite{nicer2} constrain the same quantities to be $M/M_{\odot} = 1.44 (+0.15, -0.14) $ and $R = 13.02 (+1.24, -1.06)$.  Based on NICER and XMM Newton data~\cite{PSRJ0740}, a radius of $12.71 (+1.14, -1.19)$ km is found for $M/M_{\odot} = 1.4 $. Our central value for $R_{1.4}$ falls within the lower part of these ranges.

\section{Summary and Conclusions }                                                                  
\label{Concl} 

A fully microscopic EoS up to central densities of the most massive stars---potentially involving non-nucleonic degrees of freedom and phase transitions---is not within reach. Nevertheless, neutron stars are powerful natural laboratories for constraining theories of the EoS. One must be mindful about the theory's limitations and the best ways to extract useful information from the observational constraints. Concerning the latter, there is no doubt that {\it The golden age of neutron-star physics has arrived}~\cite{M_1_2020}.

Recently, we developed EoS for NM and SNM based on high-quality 2NF at N$^3$LO and including all subleading 3NF. These were used to construct the symmetry energy and the EoS in $\beta$-equilibrated matter, from which proton and lepton fractions are easily extracted. The stellar matter EoS was then extended to densities inaccessible to chiral EFT by matching it with piecewise polytropes of different adiabatic index. From those combinations, we dropped the EoS which did not satisfy the maximum mass constraint. The causality condition is also applied.

Constraints on the radius of a medium-mass neutron star, $R_{1.4}$, are becoming more stringent, with the current uncertainty reported at about 2 km. Furthermore, $R_{1.4}$ is known to be sensitive to the pressure in neutron-rich matter near normal densities, accessible to modern effective field theories of nuclear forces.  For these reasons, we focused on predicting $R_{1.4}$ with proper uncertainty quantification.
From reports in the literature, our predicted range would increase by about 1 km on either side when additional theoretical uncertainties are included.

Based on our analysis in Section~\ref{ns_res}, we are confident that
the estimate given in Equation~(\ref{R1.4}), approximately (12 $\pm$ 1) km, is characteristic of EFT predictions based on high-quality 2NF and properly calibrated (leading and subleading) 3NF\footnote{The equations of state utilized in this work can be obtained contacting the corresponding author at the provided email address.}. The range currently cited for chiral EFT-based predictions of $R_{1.4}$ is between 10 km and 14 km, accounting for additional theoretical uncertainties.  In fact, it is interesting to notice that the extensive analysis from Ref.~\cite{LH19}, where 300,000 possible EoS were generated, provides a range for $R_{1.4}$ between 10.0 and 12.7 km, with 12.0 km being the most probable value. We recall that the outcome of PREX II gives 13.33 km as the lower limit for the radius, which is problematic to reconcile with a multitude of microscopic predictions~\cite{prexII}.

In conclusion, we reiterate that gravitational wave astronomy offers new exciting opportunities for nuclear astrophysics.  Even though chiral EFT cannot reach out to the extreme-density and yet unknown regimes at the core of these remarkable stars,
continuously improved {\it ab initio} calculations of the nuclear EoS are an essential foundation for interpreting current and future observations in terms of microscopic nuclear forces.

\section*{Acknowledgments}
This work was supported by 
the U.S. Department of Energy, Office of Science, Office of Basic Energy Sciences, under Award Number DE-FG02-03ER41270. 

%\begin{adjustwidth}{-\extralength}{0cm}

%\reftitle{References}

\end{document}